\documentclass[aps,pra,reprint,groupedaddress,superscriptaddress,floatfix]{revtex4-2}

\usepackage[T1]{fontenc}
\usepackage{newtxtext}
\usepackage{newtxmath}

\usepackage{amsmath}
\usepackage{mathtools}
\usepackage{amsfonts}

\usepackage{tikz}
\usetikzlibrary{quantikz}
\usepackage{adjustbox}

\usepackage{color}
\usepackage{xcolor}
\usepackage[colorlinks,citecolor=blue,linkcolor=blue,urlcolor=blue]{hyperref}

\usepackage{graphicx}
\usepackage[all]{hypcap} 

\begin{document}

\title{Simulating dirty bosons on a quantum computer}

\author{Lindsay Bassman Oftelie}
\thanks{These authors contributed equally to this work.}
\affiliation{Computational Research Division, Lawrence Berkeley National Laboratory, Berkeley, California 94720, USA}

\author{Roel Van Beeumen}
\affiliation{Computational Research Division, Lawrence Berkeley National Laboratory, Berkeley, California 94720, USA}

\author{Daan Camps}
\affiliation{National Energy Research Scientific Computing Center, Lawrence Berkeley National Laboratory, Berkeley, California 94720, USA}

\author{Wibe A. de Jong}
\affiliation{Computational Research Division, Lawrence Berkeley National Laboratory, Berkeley, California 94720, USA}

\author{Maxime Dupont}
\thanks{These authors contributed equally to this work.}
\affiliation{Department of Physics, University of California, Berkeley, California 94720, USA}
\affiliation{Materials Sciences Division, Lawrence Berkeley National Laboratory, Berkeley, California 94720, USA}

\begin{abstract}
    The physics of dirty bosons highlights the intriguing interplay of disorder and interactions in quantum systems, playing a central role in describing, for instance, ultracold gases in a random potential, doped quantum magnets, and amorphous superconductors. Here, we demonstrate how quantum computers can be used to elucidate the physics of dirty bosons in one and two dimensions. Specifically, we explore the disorder-induced delocalized-to-localized transition using adiabatic state preparation. In one dimension, the quantum circuits can be compressed to small enough depths for execution on currently available quantum computers. In two dimensions, the compression scheme is no longer applicable, thereby requiring the use of large-scale classical state vector simulations to emulate quantum computer performance. In addition, simulating interacting bosons via emulation of a noisy quantum computer allowed us to study the effect of quantum hardware noise on the physical properties of the simulated system. Our results suggest that scaling laws control how noise modifies observables versus its strength, the circuit depth, and the number of qubits. Moreover, we observe that noise impacts the delocalized and localized phases differently. A better understanding of how noise alters the genuine properties of the simulated system is essential for leveraging noisy intermediate-scale quantum devices for simulation of dirty bosons, and indeed for condensed matter systems in general. 
\end{abstract}

\maketitle

\section{Introduction}
Scientific investigation of dirty bosons \cite{Weichman2008,Zheludev2013} seeks to characterize the phase diagram (phases of matter, transitions) of strongly interacting bosons in a disordered environment. The combination of disorder and interactions present in these systems is a double-edged sword: one the one hand, it induces interesting quantum physics in the form of an exotic new disorder-induced phase of matter at low temperatures, namely, the Bose glass; on the other hand, it makes the study of these systems a challenging problem in condensed matter physics. Interest in dirty bosons traces back to experiments of superfluid ${}^4\mathrm{He}$ in porous media~\cite{Reppy1992,PhysRevLett.61.1954}, along with the theoretical work that followed, which identified the Bose glass phase \cite{PhysRevB.34.3136,Giamarchi1987,PhysRevB.37.325,PhysRevLett.61.1847,PhysRevB.40.546}.  While bosons typically display superfluidity and Bose-Einstein condensation in two dimensions and higher (often associated with a global quantum coherence), the Bose glass phase is characterized as gapless, compressible, short-range correlated, and localized. Therefore, upon sufficient disorder, a bosonic system can experience a delocalized-to-localized phase transition accompanied with extreme changes in its physical properties. It is analogous to the Anderson metal-to-insulator transition for noninteracting electrons~\cite{PhysRev.109.1492,RevModPhys.80.1355,abrahams2010,dobrosavljevic2012}.

The physics of dirty bosons is relevant whenever there is an interplay between disorder and interacting bosonic degrees of freedom. It has been studied in ultracold gas setups with ${}^{39}\mathrm{K}$ atoms in one-dimensional quasiperiodic optical lattice~\cite{Deissler2010,PhysRevLett.113.095301}, ${}^{6}\mathrm{Li}_2$ molecules in two dimensions~\cite{PhysRevLett.110.100601}, and ${}^{87}\mathrm{Rb}$ atoms in three dimensions~\cite{PhysRevLett.102.055301}. It has also been observed in disordered, amorphous superconductors where Cooper pairs act as effective bosonic degrees of freedom. For instance, localization of the Cooper pairs was reported in titanium nitride and indium oxide films by scanning tunneling microscope spectroscopy~\cite{PhysRevLett.101.157006,Sacepe2011}, as well as in thin films of niobium-titanium nitride and titanium nitride by microwave electrodynamics measurements~\cite{PhysRevLett.109.107003}. Furthermore, dirty boson physics play a role in chemically doped magnetic insulators~\cite{PhysRevLett.95.227201,Zheludev2013} where disorder may lead to the localization of bosonic quasiparticles. Such spin compounds include the spin-$1/2$ coupled ladders $(\mathrm{CH}_3)_2\mathrm{CHNH}_3\mathrm{Cu}(\mathrm{Cl}_{0.95}\mathrm{Br}_{0.05})_3$~\cite{PhysRevB.81.060410}, the $S=1/2$ dimer system $\mathrm{Tl}_{1-x}\mathrm{K}_x\mathrm{Cu}\mathrm{Cl}_3$~\cite{PhysRevB.83.020409}, the randomly diluted quantum spin-$1/2$ chains material $(\mathrm{Yb}_{1-x}\mathrm{Lu}_x)_4\mathrm{As}_3$~\cite{PhysRevB.94.100403}, and the $S=1$ antiferromagnet $\mathrm{Ni}(\mathrm{Cl}_{1-x}\mathrm{Br}_x)_2\mathrm{-}4\mathrm{SC}(\mathrm{NH}_2)_2$ at high magnetic fields~\cite{Yu2012,PhysRevB.86.134421,PhysRevLett.118.067203,PhysRevLett.118.067204,PhysRevB.96.024442,PhysRevLett.121.177202}. 

The problem of dirty bosons has also been considered theoretically and numerically with classical simulation methods (e.g., quantum Monte Carlo and the density matrix renormalization group)~\cite{PhysRevB.34.3136,Giamarchi1987,PhysRevB.37.325,PhysRevLett.61.1847,PhysRevB.40.546,krauth1991,scalettar1991,singh1992,sorensen1992,makivic1993,weichman1995,shang1995,pai1996,herbut1997,kisker1997,rapsch1999,alet2003,prolofev2004,giamarchi2004,priyadarshee2006,hitchcock2006,weichman2007,roux2008,gurarie2009,carrasquilla2010,altman2010,lin2011,soyler2011,ristivojevic2012,iyer2012,meier2012,alvarez2013,yao2014,ristivojevic2014,alvarez2015,ng2015,khellil2016analytical,doggen2017,PhysRevB.99.020202,dupont2019b,PhysRevB.102.174205,PhysRevE.100.030102,PhysRevE.101.042139,PhysRevE.103.052136,Takayoshi2022}. However, despite extensive investigation into dirty bosons, both experimentally and with simulation on classical computers, several fundamental questions remain open: Is there an upper critical dimension for the superfluid-to-insulator transition~\cite{PhysRevB.40.546}? Can there be an intermediate scenario between the delocalized and localized phases with a nonergodic delocalized phase similar to what is found for the Anderson localization problem on certain geometries~\cite{PhysRevB.94.220203,Biroli2018,KRAVTSOV2018148}? Is there a connection between this problem in infinite dimension and the many-body localization phenomenon~\cite{PhysRevB.102.174205,PhysRevB.100.134201,Tang2021}? Moreover, some bosonic phases of matter are hard to simulate classically due to the existence of a sign problem in quantum Monte Carlo simulations. Quantum computers, with their ability to efficiently simulate quantum systems \cite{feynman1982simulating, lloyd1996universal, abrams1997simulations, zalka1998simulating} and finely tune system parameters (e.g, disorder strength), offer a promising new platform for investigating such phases, e.g., Bose metals~\cite{PhysRevLett.106.046402,PhysRevLett.107.077201} or superglasses~\cite{PhysRevB.105.174203,PhysRevB.85.104205}. 

Already, a profusion of recent work has successfully demonstrated simulations of varied properties and behaviors of quantum systems  \cite{chiesa2019quantum, francis2020quantum, bassman2021simulating, fauseweh2021digital, sun2021quantum, guo2021observation, oftelie2022computing, chertkov2022holographic} on currently available quantum hardware. Here, we demonstrate how quantum computers can be used to simulate dirty bosons.  We focus on the toy model of hard-core bosons on a lattice in a random potential of tunable strength in one and two dimensions. We demonstrate how adiabatic state preparation \cite{aspuru2005simulated} can be used to prepare the ground state of the system on the quantum computer in order to study its properties.  In order to ensure adiabaticity, long evolution times are generally required.  Simulating this evolution generally requires deep quantum circuits, as quantum circuit depths tend to grow linearly with evolution time \cite{wiebe11, childs2018toward}. This poses a problem for current quantum computers.  This is because currently available quantum hardware is noisy, leading to short qubit decoherence times and high gate error rates, which sets an effective limit on the depth of quantum circuits that are feasible to execute \cite{Preskill2018quantumcomputingin}. 

In one dimension, we show how recently developed circuit compression techniques for free-fermionic circuits \cite{bassman2022constant, Kokcu2021, PhysRevA.105.032420, Camps2022} can be used to compress simulation of arbitrarily long time evolution of the dirty bosons into highly compact quantum circuits, thereby ensuring the adiabaticity of the evolution as well as the feasibility of successful circuit execution on current quantum hardware.  We were therefore able to obtain quantitative experimental results from IBM's quantum processors on the nature of the superfluid-to-insulator transition by extracting its critical properties such as the correlation length exponent. 

In two dimensions, the compression trick is no longer applicable and we therefore restrict our results to large-scale classical state vector simulations, which emulate quantum computation on classical computers, leveraging NVIDIA cuQuantum library on graphics processing units~\cite{cuquantum}.  Classical emulation of an ideal (i.e., noise-free) quantum computer allowed us to validate our quantum computational approach to obtaining the ground state physics of dirty bosons in two dimensions. 

Finally, we perform simulations of dirty bosons in one and two dimensions via classical emulation of a noisy quantum computer, based on a depolarizing channel of Pauli operators.  This allowed us to explore how the noise inherent in current quantum hardware affects the physical properties of the final quantum state of the simulated system. In particular, our results suggest that scaling laws control how hardware noise modifies observables versus its strength, the circuit depth, and the number of qubits. Moreover, we observe that noise impacts the delocalized and localized phases differently. We believe our results on the noise-induced physics will prove useful for simulating bosons on near-future quantum computers, which are expected to comprise more and better quality qubits.

The rest of the paper is organized as follows. In Sec. \ref{sec:method}, we define the models, physical observables of interest and the strategies used to simulate them. In Sec. \ref{sec:results}, we present and discuss our simulation results for dirty bosons in one and two dimensions, including an analysis of the observed noise-induced physics. We conclude by summarizing our findings and giving perspectives in Sec. \ref{sec:conclusion}.

\section{Model, definitions, and methods}
\label{sec:method}
\subsection{Model: Interacting bosons on a lattice}
\begin{figure}
    \includegraphics[scale=0.75]{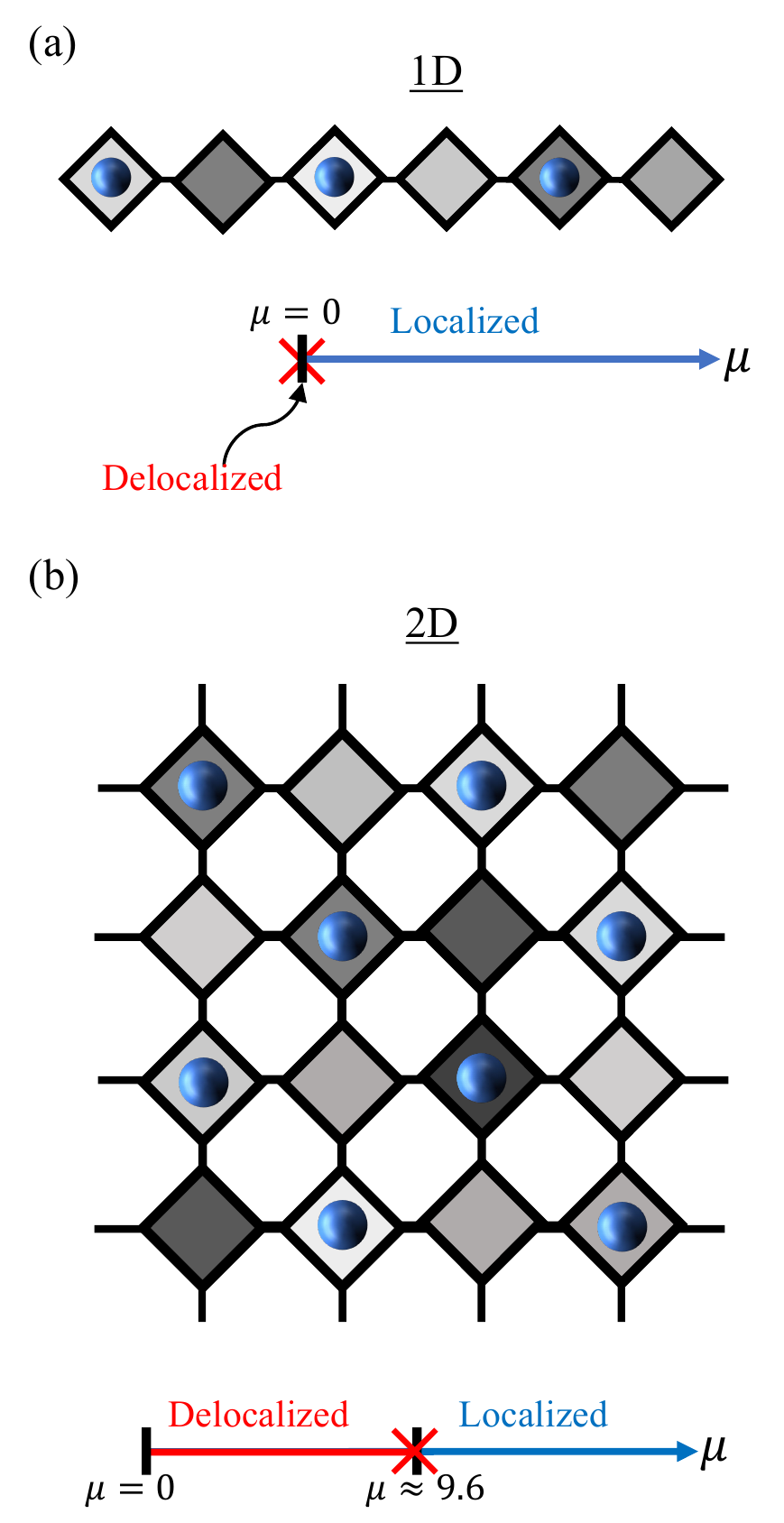} 
    \caption{Illustrations of the simulated systems along with their phase diagrams with respect to disorder strength $\mu$ in one (a) and two (b) dimensions. Diamonds represent lattice sites, whose random coloring represents their random on-site potentials.  The blue spheres represent the hard-core bosons, which are allowed to hop between lattice sites connected with an edge.  The depicted positions of bosons represents their initial configuration in all simulations.  (a) In one dimension, we simulate a chain of lattice sites with open boundary conditions, with bosons initialized at every other lattice site.  Here, we show the chain with 6 lattice sites that was simulated on IBM's quantum processor.  With no disorder ($\mu = 0$), the bosons are delocalized, but with the onset of any finite value of disorder, the bosons will localize.  (b) In two dimensions, we simulate a square lattice with periodic boundary conditions, with bosons initialized in a checkerboard pattern.  Here, we show the 16-site lattice that was simulated with the noisy quantum circuit emulator.  Up to a disorder strength of $\mu \approx 9.6$, the bosons are delocalized, but at higher disorder strength, the bosons localize.}
    \label{fig:schematic_diagrams}
\end{figure}

We consider a toy model of hard-core bosons at half-filling on a lattice, described by the following Hamiltonian,
\begin{equation}
    \hat{\mathcal{H}}_\mathrm{bosons}=-\sum\nolimits_{\langle mn\rangle}\Bigl(\hat{b}^\dag_m\hat{b}_n + \mathrm{H.c.}\Bigr)+\sum\nolimits_m\mu_m\hat{n}_m,
    \label{eq:hamiltonian_bosons}
\end{equation}
where $\hat{b}^\dag_m$ and $\hat{b}_m$ are the bosonic creation and annihilation operators on lattice site $m$, respectively. They obey the usual bosonic commutation relation $[\hat{b}_m,\hat{b}^\dag_n]=\delta_{mn}$. $\hat{n}_m=\hat{b}^\dag_m\hat{b}_n$ is the local density operator on site $m$ with the constraint $\langle\hat{n}_m\rangle\leq 1$ due to the hard-core nature of the particles. The constraint is equivalent to an infinite repulsive interaction, preventing more than one boson being on a given lattice site. Hence, the local Hilbert space of the lattice site $m$ is given by $\vert{0}_m\rangle$ and $\vert{1}_m\rangle$, corresponding to an empty and occupied lattice site $m$, respectively. The first term of Eq.~\eqref{eq:hamiltonian_bosons} describes the hopping on nearest-neighbor lattice sites. Its negative sign favors delocalization on the lattice. The second term describes a site-dependent random chemical potential drawn from a uniform distribution $\mu_m\in[-\mu,+\mu]$ with $\mu$ characterizing the strength of the disorder in the lattice. Disorder may favor localization of the particles, leading to a competition with the kinetic term. We consider two lattices: (i) a one-dimensional linear chain with open boundary conditions, shown schematically in Figure \ref{fig:schematic_diagrams}a and (ii) a two-dimensional square lattice with periodic boundary conditions, shown schematically in Figure \ref{fig:schematic_diagrams}b. We discuss the physics of each model and existing literature in their respective result section.

Up to an irrelevant constant, the Hamiltonian~\eqref{eq:hamiltonian_bosons} can be expressed in terms of Pauli operators by a Matsubara-Matsuda mapping~\cite{Matsubara1956},
\begin{equation}
    \hat{\mathcal{H}}_\mathrm{XY}=-\frac{1}{2}\sum\nolimits_{\langle{mn}\rangle}\Bigl(\hat{X}_m\hat{X}_n + \hat{Y}_m\hat{Y}_n\Bigr)+\sum\nolimits_m\frac{\mu_m}{2}\hat{Z}_m,
    \label{eq:hamiltonian_xy}
\end{equation}
equivalent to interacting spin-$1/2$ in the XY plane subject to a random spin-dependent magnetic field along the Z direction. The form of Eq.~\eqref{eq:hamiltonian_xy} directly translates into a quantum computing approach.

We consider two physical quantities to investigate the ground state properties of the dirty bosons. First, the local density,
\begin{equation}
    \bigl\langle\hat{n}_m\bigr\rangle=\frac{1}{2}\Bigl(\bigl\langle\hat{Z}_m\bigr\rangle+1\Bigr).
    \label{eq:local_density}
\end{equation}
Then the off-diagonal two-point correlation,
\begin{equation}
    C_{mn}=\Bigl\langle\hat{b}^\dag_m\hat{b}_n\Bigr\rangle=\frac{1}{2}\Bigl\langle\hat{X}_m\hat{X}_n\Bigr\rangle~~m\neq n.
    \label{eq:offdiag_correlator}
\end{equation}
Due to the random nature of the models considered, we denote disorder-averaged observables over $N_\mathrm{samples}$ realizations by $\overline{(\cdots)}$.

\subsection{Ground state preparation}

We are interested in the ground state properties of the Hamiltonian~\eqref{eq:hamiltonian_bosons} in one and two dimensions. We obtain the ground state by performing an adiabatic state preparation~\cite{aspuru2005simulated} . We start by preparing the system in a product state $\vert\Psi_\mathrm{ini}\rangle$ corresponding to a checkerboard pattern of occupied and empty lattice sites, depicted in Figure \ref{fig:schematic_diagrams}. A parent Hamiltonian to this state is,
\begin{equation}
    \hat{\mathcal{H}}_\mathrm{initial}=\sum\nolimits_m\bigl(-1\bigr)^{f(m)}\hat{Z}_m,
    \label{eq:initial_hamiltonian}
\end{equation}
with $f(m)=0$ if site $m$ is initially empty and $f(m)=1$ otherwise. We define the Hamiltonian interpolating between Eqs.~\eqref{eq:initial_hamiltonian} and~\eqref{eq:hamiltonian_xy},
\begin{equation}
    \hat{\mathcal{H}}\bigl(T,t\bigr)=\left(1-\frac{t}{T}\right)\hat{\mathcal{H}}_\mathrm{initial} + \frac{t}{T}\hat{\mathcal{H}}_\mathrm{XY},
    \label{eq:interpolating_hamiltonian}
\end{equation}
with $t\in[0,T]$. $T$ is fixed and corresponds to the total time for the unitary evolution interpolating between the two Hamiltonians (we set $\hbar=1$),
\begin{equation}
    \vert\Psi_\mathrm{fin}\rangle=\mathcal{T}\exp\left[-i\int_0^T\mathrm{d}t\;\hat{\mathcal{H}}\bigl(T,t\bigr)\right]\vert\Psi_\mathrm{ini}\rangle,
    \label{eq:final_state}
\end{equation}
where $\vert\Psi_\mathrm{fin}\rangle$ is exactly the ground state of $\hat{\mathcal{H}}_\mathrm{XY}$ in the adiabatic limit $T\to+\infty$. $\mathcal{T}$ indicates a time-ordered exponential. The evolution~\eqref{eq:final_state} conserves the number of particles in the system, which is a symmetry of the final Hamiltonian~\footnote{The conservation of the number of particles corresponds to a continuous $U(1)$ symmetry which, in any case, cannot be spontaneously broken on a finite-size system. Hence, the particle-conserving protocol will remain valid for all practical purposes, whether or not the final quantum state may spontaneously break the continuous symmetry in the thermodynamic limit.}.

In practice, the total evolution time $T$ should be chosen such that the quantum state remains in the instantaneous ground state of the interpolating Hamiltonian $\hat{\mathcal{H}}\bigl(T,t\bigr)$ throughout the time evolution. The Landau-Zener formula states that this condition is fulfilled for $T\gtrsim\Delta^{-2}_\mathrm{min}$ with $\Delta_\mathrm{min}$ the minimum spectral gap between the ground state and the first excited state along the interpolating path~\cite{RevModPhys.90.015002}. Here, because the time evolution conserves the number of particles, $\Delta_\mathrm{min}$ corresponds to the gap within this particle-conserving sector of the interpolating Hamiltonian~\footnote{From a condensed matter perspective, this gap is not necessarily the relevant one. For instance, in absence of disorder, the relevant gap in one and two dimensions is the one-particle gap as the first excited state belongs to a different sector with a number of particles $\pm 1$ with respect to half-filling.}. The choice of $T$ is discussed later following the expected physics of each model investigated.

\subsection{Implementation}

\subsubsection{Quantum circuit}

On a digital quantum computer, one needs to discretize the unitary evolution~\eqref{eq:final_state} into $N_\mathrm{steps}$ with a time step $\delta t$ such that $T=\delta tN_\mathrm{steps}$,
\begin{equation}
    \mathcal{T}\exp\left[-i\int_0^T\mathrm{d}t\;\hat{\mathcal{H}}\bigl(T,t\bigr)\right]\simeq\prod_{n=1}^{N_\mathrm{steps}}{e}^{-i\delta t\hat{\mathcal{H}}\left(\delta tN_\mathrm{steps},n\delta t\right)}.
    \label{eq:unitary_discretization}
\end{equation}
Thanks to the local nature of the interpolating Hamiltonian, Eq.~\eqref{eq:unitary_discretization} can be broken down into one- and two-site unitaries by a Suzuki-Trotter decomposition~\cite{hatano2005}. At first order, one gets,
\begin{equation}
    {e}^{-i\delta t\hat{\mathcal{H}}(T,t)}\simeq\prod\nolimits_m\hat{U}^\mathrm{Z}_m\bigl(T,t,\delta t\bigr)\prod\nolimits_{\langle{mn}\rangle}\hat{U}^\mathrm{XY}_{mn}\bigl(T,t,\delta t\bigr),
    \label{eq:unitary_trotter}
\end{equation}
with $O(\delta t^2)$ corrections not taken into account. The single-site diagonal unitary is defined as,
\begin{equation}
    \hat{U}^\mathrm{Z}_m\bigl(T,t,\delta t\bigr)=\exp\Biggl\{-i\delta t\Biggl[\Bigl(1-\frac{t}{T}\Bigr)(-1)^{f(m)}
    +\frac{t\mu_m}{2T}\Biggr]\hat{Z}_m\Biggr\},
    \label{eq:unitary_z}
\end{equation}
and the two-site off-diagonal unitary reads,
\begin{equation}
    \hat{U}^\mathrm{XY}_{mn}\bigl(T,t,\delta t\bigr)=\exp\left[i\frac{t\delta t}{2T}\Bigl(\hat{X}_m\hat{X}_n + \hat{Y}_m\hat{Y}_n\Bigr)\right].
    \label{eq:unitary_xy}
\end{equation}
The single-site unitary is equivalent to a parametric one-qubit rotation gate $\mathtt{R_Z}(\phi)=\mathrm{exp}(-i\phi\hat{Z}/2)$ and the two-site unitary to a parametric two-qubit gate $\mathtt{R_{XY}}(\phi)=\mathrm{exp}[i\phi(\hat{X}\hat{X}+\hat{Y}\hat{Y})/4]$ with the angles $\phi$ according to Eqs.~\eqref{eq:unitary_z} and~\eqref{eq:unitary_xy}. 

From the state $\vert{000\ldots{0}}\rangle$, the initial state $\vert\Psi_\mathrm{ini}\rangle$ with a checkerboard pattern of empty/occupied sites is easily obtained by applying individual $\hat{X}$ gates on the desired occupied lattice sites. For each step, the one-qubit gates of Eq.~\eqref{eq:unitary_trotter} can be applied in parallel. The two-qubit gates of Eq.~\eqref{eq:unitary_trotter} are arranged to minimize circuit depth and applied in parallel accordingly: one layer for even and odd bonds, respectively, in one dimension. On the two-dimensional square lattice, four layers are needed to accommodate the four-fold connectivity of each lattice site. The local density of Eq.~\eqref{eq:local_density} is easily computed from bitstrings obtained in the computational basis. To compute the off-diagonal two-point correlation of Eq.~\eqref{eq:offdiag_correlator}, we make a basis rotation before measurement such that the operator $\hat{X}$ is diagonal by applying Hadamard gates on the individual qubits. Expectation values are computed by performing the average over a finite set of bitstrings $N_\mathrm{shots}$.

\subsubsection{The special case of one dimension}

In one dimension, the interpolating Hamiltonian of Eq.~\eqref{eq:interpolating_hamiltonian} maps to a spinless free-fermionic system through a Jordan-Wigner transformation~\cite{Jordan1928}. Due to the free-fermionic nature of the system, the resulting quantum circuit for the state preparation can be compressed to a final depth that scales linearly with the number of qubits $N$~\cite{bassman2022constant,Kokcu2021,PhysRevA.105.032420,Camps2022}, as opposed to a depth that scales linearly with evolution, which is generally the case \cite{wiebe11, childs2018toward}. Hence, the quantum circuit depths are independent of simulation time, enabling simulation out to arbitrarily long times, ensuring adiabaticity of the evolution. Albeit smaller, the circuit keeps a similar structure after the compression, with consecutive layers of $\mathtt{R_Z}(\phi)$ and $\mathtt{R_{XY}}(\phi)$ gates and rescaled angles $\phi$ following the compression scheme. We generate compressed circuits using the open-source software package \emph{fast free fermion compiler} (\texttt{F3C}) \cite{f3c,f3cpp}, which in turn is built based on the \texttt{QCLAB} toolbox \cite{qclab,qclabpp} for creating and representing quantum circuits.

\subsubsection{Emulations on a classical computer}

Currently available quantum computers are noisy, mainly due to relatively short qubit decoherence times and relatively high gate-error rates.  This puts a limit on the size of quantum circuits that can be feasible run (i.e., return high-fidelity results) on real quantum computers, which in turn limits the system sizes that are feasible to simulate.  To gauge how simulations of larger systems may perform on improved quantum computers of the near-future, such simulations may be emulated on a classical computer. So-called \emph{quantum circuit emulation} uses a classical computer to simulate execution of quantum circuits by storing the many-body wavefunction of an $N$-qubit system (a complex vector of size $2^N$) and updating it according to the gates in the emulated quantum circuit. This will provide effective simulation results from an ideal (i.e., noise-free) quantum computer.  It is also possible to emulate a noisy quantum computer computer by adding noise channels to the emulation process, which can be tuned with a parameter that effectively sets the probability of error.  This allows us to study the effect of noise on simulation results.
Here, we use qsim~\cite{quantum_ai_team_and_collaborators_2020_4023103}, accelerate with the NVIDIA cuQuantum SDK~\cite{cuquantum}, to emulate both ideal and noisy quantum computers on a classical GPU backend. All our classical emulations were performed on the NERSC Perlmutter supercomputer.

\subsubsection{Simulations on a quantum computer}

For smaller systems sizes (around $10$ or fewer qubits), it is possible to obtain reasonable results from current quantum computers.  Here, we use IBM's `ibmq\_brooklyn' quantum processing unit (QPU), comprising qubits implemented with superconducting circuits.  While quantum circuits emulated on a classical computer may utilize arbitrary unitary gates, circuits executed on QPUs may only use so-called \emph{native} gates, a small, universal set of gates that can be natively implemented on the quantum computer. The native gate set on IBM's QPU includes the $\sqrt{\hat{X}}$ gate, which applies the square root of the Pauli-$\hat{X}$ gate; the $\mathtt{R_Z}(\phi)$ gate, which rotates the qubit about the $z$-axis by an angle $\phi$; and the two-qubit, entangling $\mathtt{CNOT}$ gate.  This is a universal gate set because any unitary gates can be decomposed into a sequence of these three native gates. Note that circuits compiled down to the native gate set can be significantly longer than their counterparts using arbitrary unitary gates. For instance, the two-qubit $\mathtt{R_{XY}}(\phi)$ gate of Eq.~\eqref{eq:unitary_xy} decomposes as,
\begin{equation}
    \begin{adjustbox}{width=0.8\columnwidth}
        \begin{quantikz}
            & \gate{\mathtt{R_X}\left(\frac{\pi}{2}\right)} & \ctrl{1} & \gate{\mathtt{R_X}\left(\frac{\phi}{2}\right)} & \ctrl{1} & \gate{\mathtt{R_X}\left(-\frac{\pi}{2}\right)} & \qw\\
            & \gate{\mathtt{R_X}\left(\frac{\pi}{2}\right)} & \targ{} & \gate{\mathtt{R_Z}\left(\frac{\phi}{2}\right)} & \targ{} & \gate{\mathtt{R_X}\left(-\frac{\pi}{2}\right)} & \qw
        \end{quantikz},
    \end{adjustbox}
    \label{eq:xy_gate_decomposition}
\end{equation}
where $\mathtt{R_X}(\phi)=\mathrm{exp}(-i\phi\hat{X}/2)$ rotates the qubit about the $x$-axis by an angle $\phi$ and which can be further decomposed using $\sqrt{\hat{X}}=e^{i\pi/4}\mathtt{R_X}(\pi/2)$ together with $\mathtt{R_Z}(\phi)$,
\begin{equation}
    \begin{adjustbox}{width=0.75\columnwidth}
        \begin{quantikz}
            & \gate{\mathtt{R_X}\bigr(\phi\bigl)} & \qw
        \end{quantikz}
        \begin{quantikz}=\end{quantikz}
        \begin{quantikz}
            & \gate{\mathtt{H}} & \gate{\mathtt{R_Z}\bigr(\phi\bigl)} & \gate{\mathtt{H}} & \qw
        \end{quantikz},
    \end{adjustbox}
    \label{eq:rx_gate_decomposition}
\end{equation}
with
\begin{equation}
    \begin{adjustbox}{width=0.85\columnwidth}
        \begin{quantikz}
            & \gate{\mathtt{H}} & \qw
        \end{quantikz}
        \begin{quantikz}=\end{quantikz}
        \begin{quantikz}
            & \gate{\mathtt{R_Z}\left(\frac{\pi}{2}\right)}  & \gate{\mathtt{R_X}\left(\frac{\pi}{2}\right)} & \gate{\mathtt{R_Z}\left(\frac{\pi}{2}\right)} & \qw
        \end{quantikz},
    \end{adjustbox}
    \label{eq:h_gate_decomposition}
\end{equation}
the Hadamard gate. Global phases are irrelevant.

\section{Results}
\label{sec:results}

\subsection{Dirty bosons in one dimension}
\label{sec:bosons_1d}

\subsubsection{Review of the existing literature and expectations}

In one dimension, the bosonic model of Eq.~\eqref{eq:hamiltonian_bosons} can be mapped to a spinless fermionic model with local terms by a Jordan-Wigner transformation~\cite{Jordan1928},
\begin{equation}
    \hat{\mathcal{H}}_\mathrm{fermions}=-\sum\nolimits_{\langle{mn}\rangle}\Bigl(\hat{c}^\dag_m\hat{c}_n + \mathrm{H.c.}\Bigr)+\sum\nolimits_m\mu_m\hat{n}_m,
    \label{eq:hamiltonian_fermions}
\end{equation}
where $\hat{c}^\dag_m$ and $\hat{c}_m$ are the fermionic creation and annihilation operators on lattice site $m$, respectively~\footnote{For periodic boundary conditions, there is an additional boundary term in Eq.~\eqref{eq:hamiltonian_fermions}, but we only consider open boundary conditions  for the one-dimensional case in this work.}. They obey the usual fermionic anticommutation relation $\{\hat{c}_m,\hat{c}^\dag_n\}=\delta_{mn}$. $\hat{n}_m=\hat{c}^\dag_m\hat{c}_n$ is the local density operator on site $m$. Unlike the hard-core bosons, effectively interacting through an infinite repulsive interaction, the fermions are non-interacting in Eq.~\eqref{eq:hamiltonian_fermions}. The Hamiltonian is quadratic, and its eigenstates are easily constructed by filling one-particle orbitals. In the absence of a disordered potential ($\mu=0$), the system displays superfluidity and quasi-long-range order with critical properties captured by Tomonaga-Luttinger liquid physics~\cite{giamarchi2004}. For any nonzero disorder strength $\mu>0$, all the single-particle orbitals will experience Anderson localization, and therefore, all the eigenstates of Eq.~\eqref{eq:hamiltonian_fermions} will also be localized~\cite{Giamarchi1987,PhysRevB.37.325}, including its ground state. The localization is characterized by a localization length scale $\xi$ diverging as $\xi\sim\mu^{-2}$ as $\mu\to 0$~\cite{Thouless1972}, such that the critical behavior is recovered at $\mu=0$. In the other limit, at strong disorder, the localization length decreases as $\xi\sim(\ln\mu)^{-1}$~\cite{Laflorencie2021}.

Hence, a localized phase exists for $\mu>0$ and delocalization only at $\mu=0$, which is also the transition point. The phase diagram is depicted in Figure \ref{fig:schematic_diagrams}a.  In order to stabilize a delocalized phase for hard-core bosons at finite $\mu$, one can add a nearest-neighbor interaction term of the form $2U\sum_{\langle{mn}\rangle}\hat{n}_m\hat{n}_n$ in the Hamiltonian, with strength $U$. If the interaction is attractive and in the range $U\in(-1,-1/2)$, such that disorder is irrelevant in a renormalization group sense, delocalized Tomonaga-Luttinger liquid physics can be robust up to $\mu\lesssim 0.74$~\cite{Giamarchi1987,PhysRevB.37.325,doggen2017}. Note that in that case, the localization length diverges differently as the delocalized phase is approached, with $\xi\sim\mu^{2/(2K-3)}$ where $K=\pi/2\arccos(-U)$ the so-called Tomonaga-Luttinger liquid parameter~\cite{PhysRevB.12.3908}---for the case $U=0$ considered in this work we have $K=1$.

\subsubsection{Gap scaling and state preparation protocol}

For the state preparation to be accurate, it needs to be adiabatic, which requires prior knowledge of the minimum spectral gap $\Delta_\mathrm{min}$ between the instantaneous ground state and the first excited state during the evolution protocol. One can then choose a total evolution time $T\gtrsim\Delta^{-2}_\mathrm{min}$ according to the Landau-Zener formula.

\begin{figure}[t]
    \includegraphics[width=1\columnwidth]{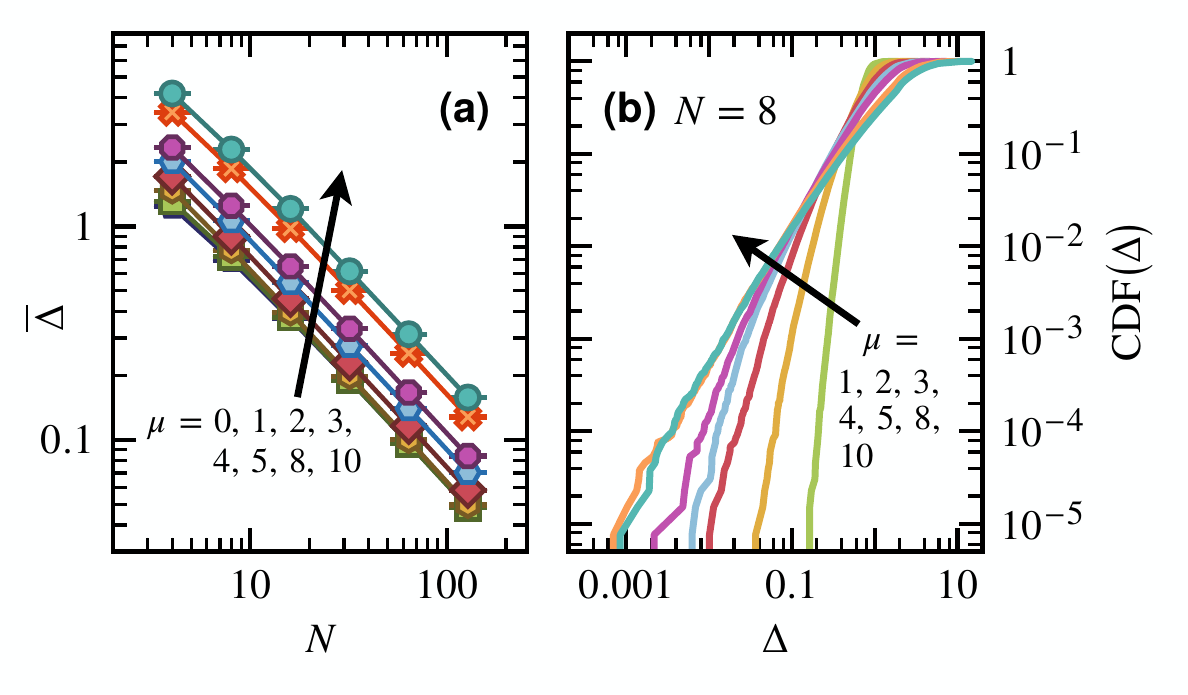} 
    \caption{Exact diagonalization results for the one-dimensional case of Eq.~\eqref{eq:hamiltonian_bosons}. The data include $N_\mathrm{samples}=2^{17}$ random disorder configurations. (a) Average spectral gap $\overline{\Delta}$ between the ground state and the first excited state in the half-filling sector as a function of the system size $N$ for different disorder strengths $\mu$. We find that $\overline{\Delta}\sim N^{-1}$. (b) Cumulative distribution function of the spectral gap $\Delta$ for various disorder strengths $\mu$ at fixed system size $N=8$.}
    \label{fig:gap_one_dimension}
\end{figure}

Instead of $\Delta_\mathrm{min}$ which is cumbersome to evaluate, we consider the gap $\Delta$ of the final state. We perform exact classical simulations diagonalizing the final Hamiltonian. We show in Fig.~\ref{fig:gap_one_dimension}(a) the average gap computed over more than $N_\mathrm{samples}=10^5$ random disorder realizations versus the system size for various disorder strengths $\mu$. In absence of disorder, the gap follows $\Delta\sim N^{-1}$ with $N$ the system size, which is compatible with the critical properties of the Tomonaga-Luttinger phase. We find the same scaling in presence of disorder, see Fig.~\ref{fig:gap_one_dimension}(a). However, the gap is widely distributed over several orders of magnitude. We illustrate this for $N=8$ with the cumulative distribution function of the gap $\mathrm{CDF}(\Delta)$ in Fig.~\ref{fig:gap_one_dimension}(b). We observe that as the disorder strength increases, there exist random realizations with smaller and smaller gaps.

In the following, we decide to take a total evolution time $T=10^4$ (along with a time step $\delta{t}=0.05$), which will lead to an adiabatic state preparation for gaps $\Delta_\mathrm{min}\gtrsim 10^{-2}$. For $N=8$ and the disorder strengths $\mu\leq{10}$, Fig.~\ref{fig:gap_one_dimension}(b) states that the probability to find such a disorder configuration with a gap smaller than $\approx{10}^{-2}$ is less than $\approx{0.1}\%$. Hence, at most, a disordered sample for every one thousand will not be prepared adiabatically.

\subsubsection{Local density of particles}

\begin{figure}[t]
    \includegraphics[width=1\columnwidth]{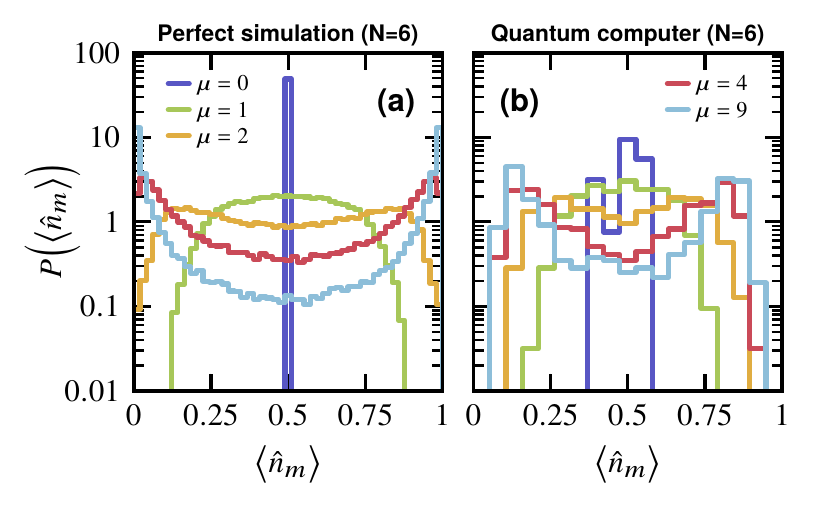} 
    \caption{Probability distribution for $N=6$ of the local density of particles of Eq.~\eqref{eq:local_density} for various disorder strengths. (a) Exact diagonalization data computed over $N_\mathrm{samples}=2^{13}$ random disorder realizations. (b) Experimental data from the quantum computer over a hundred random disorder realizations for each disorder strength with $N_\mathrm{shots}=2^{13}$ measurements for each realization.}
    \label{fig:local_density_one_dimension}
\end{figure}

We first consider the local density of particles of Eq.~\eqref{eq:local_density}. For each disordered sample of length $N$, we collect $N$ expectation values and compute the probability distribution $P(\langle\hat{n}_m\rangle)$ with $\langle\hat{n}_m\rangle\in[0,1]$. The quantity displays no size effect (data not shown) and we consider $N=6$. Exact diagonalization data is plotted in Fig.~\ref{fig:local_density_one_dimension}(a). In absence of disorder, the system being at half-filling is reflected by a delta-peaked distribution at $\langle\hat{n}_m\rangle={1/2}$. As the disorder strength increases, the distribution spreads out until at larger disorder the distribution takes a U-shape with the maximum probability for $\langle\hat{n}_m\rangle\to 0,~1$. At strong disorder, the bosons will be predominantly present on lattice sites with a negative chemical potential $\mu_m<0$ to minimize the energy of the system, in line with a localization picture. In fact, it has been found to lead in the thermodynamic limit to a chain breaking mechanism as $N\to+\infty$~\cite{PhysRevB.100.134201}, holding in presence of nearest-neighbor interaction and for excited states in the context of the many-body localization phenomenon~\cite{PhysRevResearch.2.042033}.

Using the same procedure, we plot the $N=6$ qubits experimental probability distribution of the local density of particles in Fig.~\ref{fig:local_density_one_dimension}(b). We considered a hundred random realizations for each disorder strength. For each disordered sample, the expectation value of the local density is computed from $N_\mathrm{shots}=2^{13}$ measurements according to Eq.~\eqref{eq:local_density}. The general features of the exact data are reproduced experimentally: From small to strong disorder, the maximum shifts from half-filling to the extrema values $\langle\hat{n}_m\rangle\to 0,~1$ to form a U-shaped distribution---though the U shape doesn't extend all the way to $\langle\hat{n}_m\rangle\to 0,~1$. As we will see in Sec.~\ref{sec:noise_physics}, the discrepancy is due to inherent hardware noise which favors delocalization.

\subsubsection{Delocalization-localization transition}

\begin{figure*}[t]
    \includegraphics[width=2\columnwidth]{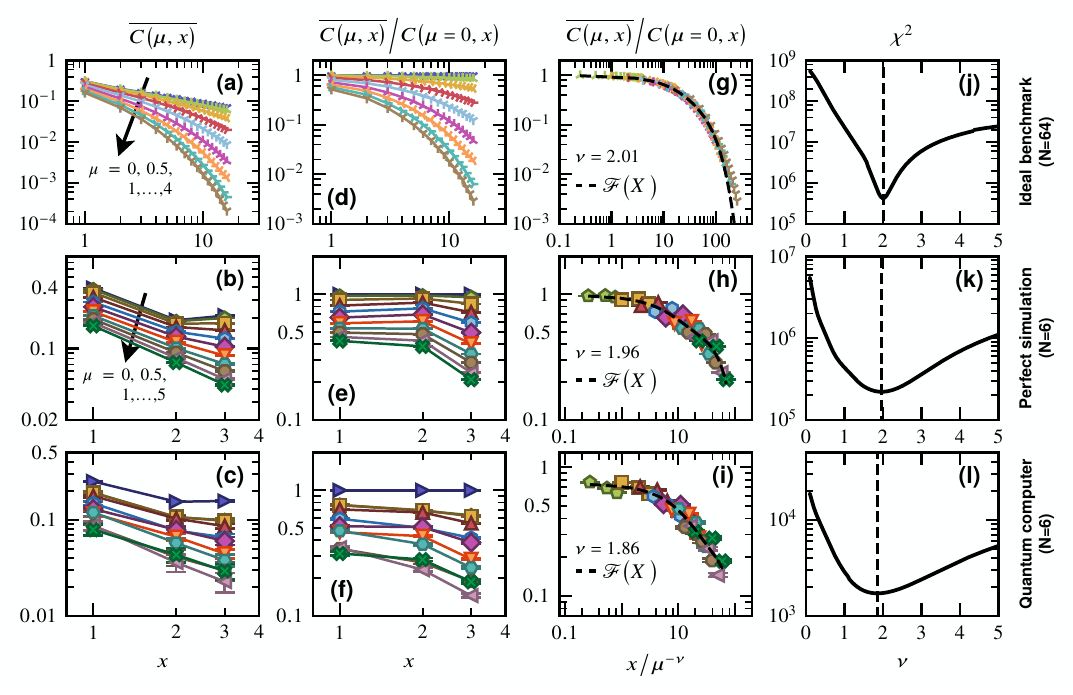} 
    \caption{First row: Ideal benchmark based on exact diagonalization for $N=64$ averaged over $N_\mathrm{samples}=2^{13}$ random realizations. Data points close to the boundaries are dismissed. Error bars are smaller than the symbols and not shown. Second row: Perfect simulation based on exact diagonalization for $N=6$ averaged over $N_\mathrm{samples}=2^{13}$ random realizations. Third row: Quantum computer result for $N=6$ averaged over a hundred random configurations with $N_\mathrm{shots}=2^{13}$ measurements each. The legend is the same as in the second row. First column: Average correlator versus the distance for different disorder strengths $\mu$. Second column: Average correlator rescaled by the disorder-free data versus the distance for different disorder strengths $\mu$. Third column: Average rescaled correlator versus the rescaled distance using the exponent $\nu$ minimizing the chi-square in the fourth column. Fourth column: Chi-square $\chi^2$ characterizing the quality of the data collapse as a function of the correlation length exponent $\nu$ (smaller is better). The vertical dashed line is positioned at the minimum $\chi^2$ value for which $\nu$ leads to the best data collapse ($\nu=2$ is the expected theoretical value).}
    \label{fig:critical_scaling_one_dimension}
\end{figure*}

We now turn our attention to the two-point off-diagonal correlator of Eq.~\eqref{eq:offdiag_correlator}. For a system of length $N$, we take the reference lattice site to be $N/2$ and note the distance with the other lattice site by $x$. We plot in Figs.~\ref{fig:critical_scaling_one_dimension}(a),~\ref{fig:critical_scaling_one_dimension}(b), and~\ref{fig:critical_scaling_one_dimension}(c) the correlator versus the distance $x$ for various disorder strengths $\mu$ for three cases. The first case is from an ideal benchmark based on exact diagonalization data on $N=64$. The second case is from a perfect simulation on $N=6$ based on exact diagonalization. The third case is the experimental data on $N=6$ from the quantum computer.

In absence of disorder, the system is critical and displays a power-law decay of the form $C(\mu=0,x)\sim{1/\sqrt{x}}$ for $N\to+\infty$ and $x\to+\infty$. The expectation is that, close to the transition, the disorder will modify this scaling by introducing a length scale $\xi$,
\begin{equation}
    \overline{C\bigl(\mu,x\bigr)}=C\bigl(\mu=0,x\bigr)\times\mathcal{F}\bigl(x/\xi\bigr),
    \label{eq:critical_scaling_correlator}
\end{equation}
To isolate the scaling function $\mathcal{F}(X)$, we start by plotting in Figs.~\ref{fig:critical_scaling_one_dimension}(d),~\ref{fig:critical_scaling_one_dimension}(e), and~\ref{fig:critical_scaling_one_dimension}(f) the correlator divided by its disorder-free value. For a second order transition, the length scale will diverge algebraically as the critical point is approached, i.e., $\xi\sim\mu^{-\nu}$ with $\nu$ the correlation length critical exponent. Hence, plotting the data as a function of the variable $x\mu^\nu$ should lead to a collapse onto a single function $\mathcal{F}(X)$. The goal is now to find the value of $\nu$ leading to the best data collapse. We employ the procedure of Ref.~\onlinecite{PhysRevB.106.L041109} which Taylor expands the unknown scaling function,
\begin{equation}
    Y=\mathcal{F}\bigl[X(\nu)\bigr]\approx\Bigl[\sum\nolimits_{m=0}^Ma_mX^m(\nu)\Bigr]\times{e}^{-\tilde{a}X(\nu)},
    \label{eq:taylor_expansion_scaling_func}
\end{equation}
with an extra exponential term accounting for the rapid decay observed in Fig.~\ref{fig:critical_scaling_one_dimension}(d) that would otherwise require a high expansion order $M$ to be captured in practice. In Eq.~\eqref{eq:taylor_expansion_scaling_func}, $Y\equiv \overline{C(\mu,x)}/C(\mu=0,x)$ and $X(\nu)\equiv x\mu^\nu$. For a given value of the exponent $\nu$, we perform a least-square fitting of the data with parameters $\tilde{a}$, $a_0$, $a_1$, etc. The quality of the fit for $K$ pairs of data points $\{Y_k,X_k\}$ is measured by the chi-square,
\begin{equation}
    \chi^2=\sum\nolimits_{k=1}^K\left(\frac{\mathcal{F}\bigl[X_k(\nu)\bigr]-Y_k}{\Delta Y_k}\right)^2,
    \label{eq:chi_square}
\end{equation}
with $\Delta Y_k$ the error on the data point $Y_k$. We repeat the procedure for a grid of the exponent $\nu$ and find the value leading to the best data collapse as the one minimizing $\chi^2$ in Eq.~\eqref{eq:chi_square}. We use $M=3$ in practice and plot the results in Figs.~\ref{fig:critical_scaling_one_dimension}(j),~\ref{fig:critical_scaling_one_dimension}(k), and~\ref{fig:critical_scaling_one_dimension}(l). The value of $\nu$ minimizing $\chi^2$ is then used in Figs.~\ref{fig:critical_scaling_one_dimension}(g),~\ref{fig:critical_scaling_one_dimension}(h), and~\ref{fig:critical_scaling_one_dimension}(i) to rescale accordingly the $x$ axis. We theoretically expect $\nu=2$ and find that the value is perfectly recovered in the ideal benchmark case~\cite{Thouless1972,Giamarchi1987,PhysRevB.37.325}. Although less satisfying, we can obtain a data collapse for the perfect simulation on $N=6$ with an exponent $\nu\approx{1.96}$.

The procedure that was carried on the ideal and perfect data is repeated on the experimental data, see Figs.~\ref{fig:critical_scaling_one_dimension}(c),~\ref{fig:critical_scaling_one_dimension}(f),~\ref{fig:critical_scaling_one_dimension}(i), and~\ref{fig:critical_scaling_one_dimension}(l). Each data point corresponds to an average over a hundred disordered realizations, where the correlator of Eq.~\eqref{eq:offdiag_correlator} has been computed from $N_\mathrm{shots}=2^{13}$ measurements. As expected, we observe a decay of the correlator with the distance and the hierarchy of the different curves with respect to the disorder strength is broadly respected when compared to the perfect simulation results. Proceeding with the data collapse, we find that the chi-square is minimized for $\nu\approx 1.86$. The resulting scaling curve is qualitatively similar to the perfect simulation one. The difference is attributed to the inherent hardware noise. We study its effect on the physics of dirty bosons in Sec.~\ref{sec:noise_physics}. The correlation length exponent $\nu\approx 1.86$ is close to the expected theoretical value of $\nu=2$.

\subsection{Dirty bosons in two dimensions}
\label{sec:bosons_2d}

\subsubsection{Review of the existing literature and expectations}

In two dimensions, the physics of the model~\eqref{eq:hamiltonian_bosons} is different than its one-dimensional counterpart. In the thermodynamic limit, the larger dimensionality allows for a spontaneous breaking of the continuous $U(1)$ symmetry of the model. Up to a critical disorder strength $\mu_\mathrm{c}\lesssim 9.6$~\cite{alvarez2015,ng2015}, the symmetry is spontaneously broken and the bosons form a Bose-Einstein condensate with true long-range order, unlike the one-dimensional case which was limited to quasi-long-range order. We note that on a finite system, $U(1)$ cannot be spontaneously broken---though one can engineer quantum states breaking it explicitly~\cite{DallaTorre2022}. At larger disorder strength, the system enters a Bose glass phase where condensation and superfluidity are destroyed and localization takes place. The phase diagram is depicted in Figure \ref{fig:schematic_diagrams}b.  The transition between the two phases is second order, with critical exponents that have been estimated by classical quantum Monte Carlo simulations~\cite{makivic1993,shang1995,priyadarshee2006,alvarez2013,ng2015}. A peculiarity of the superfluid-to-insulator transition is the conjecture that the dynamical exponent $z$ equals the dimensionality $d$ of the model~\cite{PhysRevB.40.546}, which is in line with quantum Monte Carlo estimates in $d=1$, $2$, and $3$.

\subsubsection{Delocalized versus localized bosons}

\begin{figure}[t]
    \includegraphics[width=1\columnwidth]{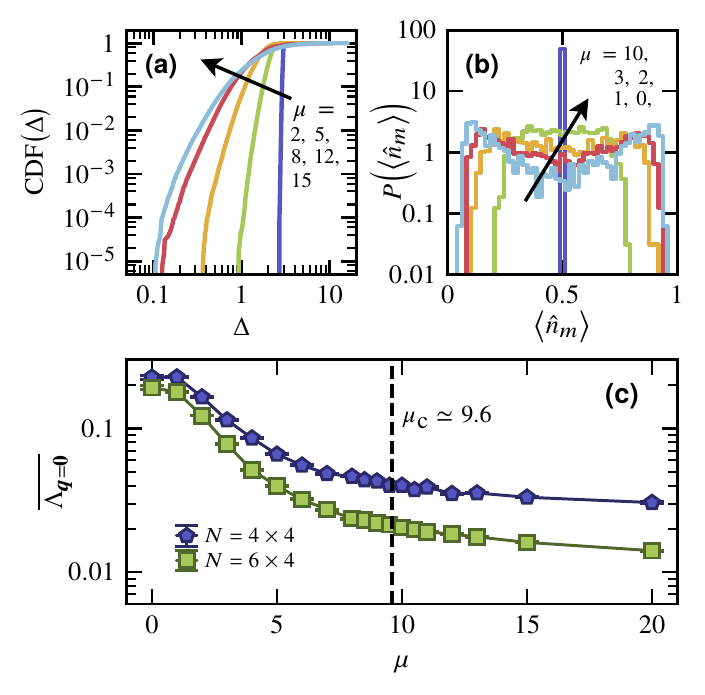} 
    \caption{(a) Cumulative distribution function of the spectral gap $\Delta$ between the ground state and first excited state in the half-filling sector. We consider a fixed system size $N=4\times 4$ at half-filling for various disorder strengths $\mu$. Data obtained by exact diagonalization by averaging over $N_\mathrm{samples}=2^{17}$ random disordered samples. (b) Probability distribution for $N=4\times 4$ of the local density of particles of Eq.~\eqref{eq:local_density} for various disorder strengths. Data computed from quantum circuit for adiabatic state preparation over a hundred random disorder realizations. (c) Average coherent density of bosons $\overline{\Lambda_{\boldsymbol{q}=\boldsymbol{0}}}$ of Eq.~\eqref{eq:bec_density} for $N=4\times 4$ as a function of the disorder strength $\mu$. Each data point is averaged over a hundred random realizations whose ground state is obtained by an adiabatic state preparation circuit. From quantum Monte Carlo simulations, we expect the transition between the superfluid and insulating phases to take place at $\mu_\mathrm{c}\approx 9.6$~\cite{alvarez2015,ng2015}.}
    \label{fig:bosons_2d}
\end{figure}

\paragraph{Scaling of the gap---} To choose the correct total time evolution for the protocol to be adiabatic, we consider the spectral gap $\Delta$ between the ground and first excited states in the half-filling sector of a $N=4\times 4$ system~\footnote{From a condensed matter perspective, the relevant gap in the symmetry-breaking phase ($\mu\leq\mu_\mathrm{c}$) is the one-particle gap between the ground state in the half-filling sector and the ground state in the sector at half-filling plus or minus one particle. As we expect an Anderson tower of states~\cite{PhysRev.86.694,Lhuillier2005}---though most commonly considered for the spontaneous breaking of an $SU(2)$ symmetry---, the one-particle gap scaling as $\sim 1/N$ will act as a lower bound for $\Delta$ defined within the half-filling sector.}. We perform exact diagonalization calculations on the final Hamiltonian for various disorder strengths $\mu$ and plot the cumulative distribution function $\mathrm{CDF}(\Delta)$ in Fig.~\ref{fig:bosons_2d}(a) using over $N_\mathrm{samples}=10^5$ random realizations. As the disorder strength increases, there exist random realizations with smaller and smaller gaps. For $\mu=15$ (one of the largest disorder strengths considered), we find that the minimum gap is $\approx 0.1$, which requires a total time evolution $T=100$ for adiabaticity. We use this value along with a time step $\delta t=0.005$ in the following.

\paragraph{Local density of bosons---} We now perform simulations using a state preparation protocol based on a quantum circuit for $N=4\times 4$. At the end of the circuit, we compute the local density of bosons of Eq.~\eqref{eq:local_density}. We show their distribution in Fig.~\ref{fig:bosons_2d}(b) for various disorder strengths $\mu$. In absence of disorder ($\mu=0$), all individual lattice sites are at half-filling. As the disorder strength increases, the distribution broadens toward a U shape, similar to the one-dimensional case studied in Fig.~\ref{fig:local_density_one_dimension}.

\paragraph{Coherent density of bosons---} We consider the momentum distribution function based on the two-point off-diagonal correlation function of Eq.~\eqref{eq:offdiag_correlator},
\begin{equation}
    \Lambda_{\boldsymbol{q}}=\frac{1}{N(N-1)}\sum\nolimits_{m\neq n}\cos\bigl(\boldsymbol{q}\cdot\boldsymbol{r}_{mn}\bigr)\Bigl\langle\hat{b}^\dag_m\hat{b}_n\Bigr\rangle,
    \label{eq:bec_density}
\end{equation}
with $\boldsymbol{q}$ the momentum and $\boldsymbol{r}_{mn}$ the distance between lattice sites $m$ and $n$~\footnote{At half-filling, taking into account the diagonal elements $m=n$ leads to a contribution $N/2$ from the sum. Adapting the normalization to $1/N^2$ because of the new terms, the normalized contribution reads $1/2N$, which goes to zero in the thermodynamic limit.}. In translationally invariant systems, Bose-Einstein condensation (BEC) typically occurs when a macroscopic fraction of bosons occupies the single momentum component $\boldsymbol{q}=\boldsymbol{0}$, and $\Lambda_{\boldsymbol{q}=\boldsymbol{0}}$ plays the role of the order parameter for the symmetry-breaking BEC phase. In systems that are not translation invariant (e.g., because of disorder) some bosons may condense in other momentum components $\boldsymbol{q}\neq\boldsymbol{0}$, and it is more rigorous to consider the one-body density matrix together with the Penrose-Onsager criterion~\cite{penrose1951, penrose1956,yang1962} to assess condensation. Nevertheless, despite subtle differences (see, e.g., Ref.~\onlinecite{PhysRevA.91.023602} and references therein), $\Lambda_{\boldsymbol{q}=\boldsymbol{0}}$, known as the coherent density, has been found to provide accurate information for the problem of interest in two dimensions~\cite{alvarez2015,ng2015}.

For $N=4\times 4$ and $N=6\times 4$, we compute the coherent density as a function of the disorder strength. We plot the results in Fig.~\ref{fig:bosons_2d}(c), where one observes that at fixed system size, it decays as the disorder strength increases. As $N\to\infty$, one expects $\Lambda_{\boldsymbol{q}=\boldsymbol{0}}(\mu<\mu_\mathrm{c})\to\mathrm{constant}$ and $\Lambda_{\boldsymbol{q}=\boldsymbol{0}}(\mu\geq\mu_\mathrm{c})\to 0$ in the delocalized and localized phases, respectively. With only two system sizes, we cannot perform a finite-size scaling analysis. We note that at criticality ($\mu=\mu_\mathrm{c}$), a system of linear length $L$ will display a behavior of the form $\Lambda_{\boldsymbol{q}=\boldsymbol{0}}(\mu=\mu_\mathrm{c})\sim L^{-z-\eta}$, with $z$ and $\eta$ the dynamical ab anomalous exponents~\cite{alvarez2015,ng2015}, respectively.

\subsection{Noise-induced physics}
\label{sec:noise_physics}

\subsubsection{Noise model}

We now investigate the effect of the noise on the physics of dirty bosons. To that end, we consider a stochastic Pauli error model with quantum channels $\mathcal{E}_{m}$ and $\mathcal{E}_{mn}$ applied after each one- and two-qubit gates, respectively~\cite{Nielsen2011},
\begin{equation}
    \begin{aligned}
        \mathcal{E}_m\bigl(\hat{\rho}\bigr)=\bigl(1-p\bigr)\hat{\rho} + \frac{p}{3}\sum\nolimits_{a}\hat{\sigma}_m^a\hat{\rho}\hat{\sigma}_m^a,~~\mathrm{and}\\
        \mathcal{E}_{mn}\bigl(\hat{\rho}\bigr)=\bigl(1-p\bigr)\hat{\rho} + \frac{p}{15}\sum_{(a,b)\neq(0,0)}\hat{\sigma}_m^a\hat{\sigma}_n^b\hat{\rho}\hat{\sigma}_m^a\hat{\sigma}_n^b,
    \end{aligned}
    \label{eq:pauli_channels}
\end{equation}
with $\hat{\rho}$ the density matrix of the system, $\hat{\sigma}^0_m\equiv\hat{I}_m$,  $\hat{\sigma}^1_m\equiv\hat{X}_m$, $\hat{\sigma}^2_m\equiv\hat{Y}_m$, and $\hat{\sigma}^3_m\equiv\hat{Z}_m$ are Pauli matrices. $p\in[0,1]$ is a parameter controlling the error rate, which for simplicity is taken the same for all qubits and gates. In practice, the parameter $p$ in Eq.~\eqref{eq:pauli_channels} can be made dependent of the qubits ($m$, $n$) and indices ($a$, $b$) and fit a specific quantum device through a noise reconstruction protocol~\cite{Erhard2019,Flammia2020}. This noise model has been been considered successfully in other condensed matter studies~\cite{PRXQuantum.2.030346,PhysRevLett.126.230501,PhysRevB.106.L041109,Azses2022}. In particular, the noise-depth tradeoff in adiabatic quantum circuits was studied in Ref.~\onlinecite{Azses2022}---here, the depth will be kept fixed.

We simulate the noise model of Eq.~\eqref{eq:pauli_channels} using ``quantum trajectories''~\cite{PhysRevLett.52.1657,PhysRevLett.68.580}, which allow to work with pure states instead of the density matrix $\hat{\rho}$. Many instances of the same circuit are simulated and each one of them is stochastically subject to the addition of random Pauli gates with a probability $\propto p$ following Eq.~\eqref{eq:pauli_channels}. The probability that no Pauli gate is added after a one- or two-qubit gate is $(1-p)$. Here, each shot (output bitstring) is a different stochastic random Pauli realization.

In one dimension, we simulate systems up to $N=20$ using a total time evolution $T=5\times 10^4$ along with a time step $\delta t=0.05$, i.e., a total of $10^6$ time steps compressed into a circuit depth $O(N)$. For each disorder strength, the data is averaged over $N_\mathrm{samples}=2^{10}$ random realizations, and observables are computed from $N_\mathrm{shots}=2^{10}$ measurements, each corresponding to a different quantum trajectory. In two dimensions, we simulate a system size of $N=4\times 4$ with periodic boundary conditions. We perform a total time evolution $T=10^2$ using a time step $\delta t=0.025$, corresponding to a total of $4,000$ steps. The data is averaged over $N_\mathrm{samples}=10^2$ random realizations and observables are computed from $N_\mathrm{shots}=10^2$ measurements.

\subsubsection{Noise-induced modifications on the physical properties}

It was argued in Ref.~\onlinecite{PhysRevB.106.L041109} that a noise model such as Eq.~\eqref{eq:pauli_channels} on an adiabatic state preparation protocol was to introduce a length scale $\xi_\mathrm{noise}\sim 1/pD$ in the system where $D$ is the circuit depth and $p$ the noise strength. For instance, a two-point correlation function probing critical properties with two local operators separated by a distance $x>0$ in real space would be modified by the noise as $C(x,p)\sim C(x,p=0)\times\exp(-x/\xi_\mathrm{noise})$~\footnote{Here, probing the critical properties implies that the noiseless correlator $C(x,p=0)$ decays algebraically with the distance $x$.}. Another interpretation of the result is that the average number of random Pauli operators in the circuit volume $Dx$ is $pDx$. Each of them will reduce the correlation by a positive multiplicative factor $\epsilon<1$, amounting for a total $C(x,p)\sim C(x,p=0)\times\epsilon^{pDx}$ in the corresponding circuit volume. This is equivalent to the above result with $\xi_\mathrm{noise}=-(pD\ln\epsilon)^{-1}$. Along the same lines, for a local observable $O$ defined such that it takes a finite value in a perfect simulation and zero for a random state (broadly defined as what an excessive amount of noise would lead to), one expects that $O(p)\sim O(p=0)\times\exp(-\tilde{\epsilon}pD)$ with $\tilde{\epsilon}\equiv-\ln\epsilon>0$. We now seek to characterize the effect of the noise on the physical properties and see how it fits with previous findings.

\subsubsection{The local density of particles}

\begin{figure}[t]
    \includegraphics[width=1\columnwidth]{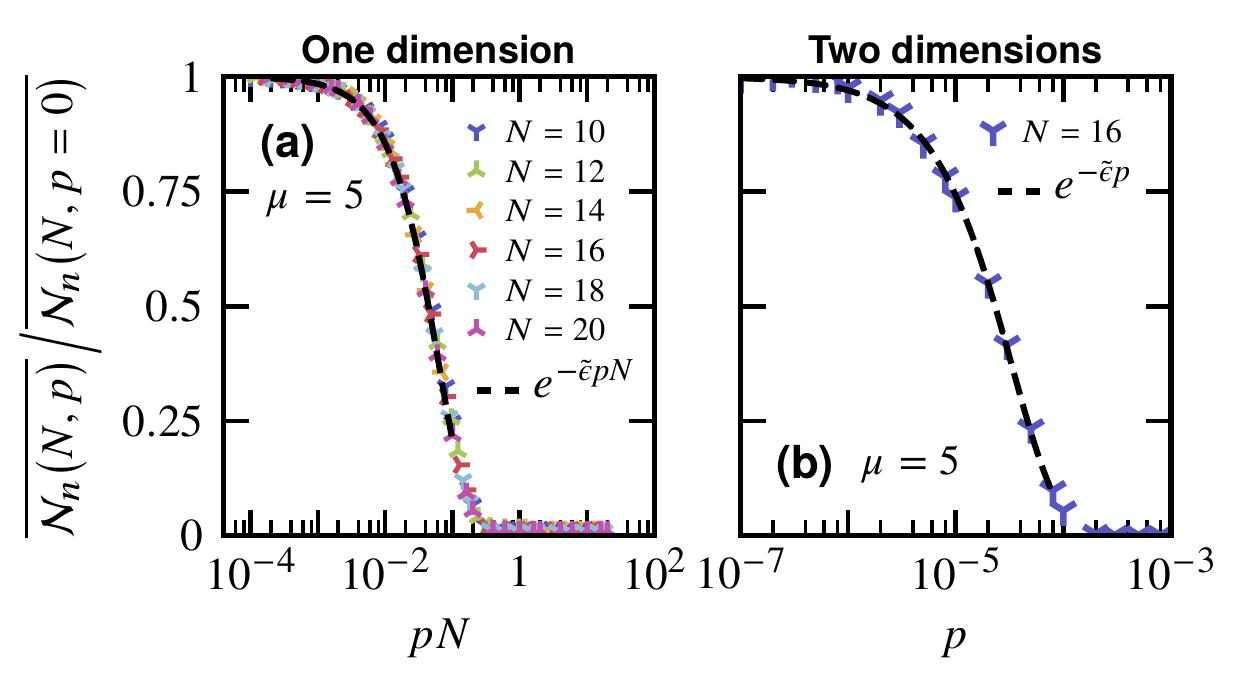} 
    \caption{Quantity of Eq.~\eqref{eq:perfect_occ} based on the local density of particles at fixed disorder strength $\mu$ for different noise strengths $p$ and system sizes $N$. (a) One-dimensional data for $\mu=5$ based on $N_\mathrm{samples}=2^{10}$ independent disordered samples and $N_\mathrm{shots}=2^{10}$ random noise measurements for each. Error bars are smaller than the symbols and not shown. For $pN\ll 1$, we observe a behavior of the form $\sim\exp(-\tilde{\epsilon}pN)$ with $\tilde{\epsilon}=15.7(2)$ a fitting parameter. Here, $N$ has to be interpreted as the depth $D$ of the circuit, which scales linearly with $N$ after the free-fermionic compression (b) Two-dimensional data for $\mu=5$ based on a hundred independent disordered samples and $N_\mathrm{shots}=10^2$ random noise measurements for each. For $p\to 0$, we observe an exponential decay $\sim\exp(-\tilde{\epsilon}p)$ with $\tilde{\epsilon}=30(2)\times 10^3$ a fitting parameter.}
    \label{fig:noisy_physics_local_density}
\end{figure}

First, we consider a quantity based on the local density of Eq.~\eqref{eq:local_density},
\begin{equation}
    \mathcal{N}_m=\left\vert{2\bigl\langle\hat{n}_m\bigr\rangle - 1}\right\vert\in [0,1],
    \label{eq:perfect_occ}
\end{equation}
which measures how the lattice site $m$ is close to perfect (non)occupation. One gets $\mathcal{N}_m\to 1$ for perfect (non)occupation $\bigl\langle\hat{n}_m\bigr\rangle\to 0$ or $1$, while $\mathcal{N}_m=0$ for $\bigl\langle\hat{n}_m\bigr\rangle=1/2$. In absence of disorder ($\mu=0$), one expects $\mathcal{N}_m=0~\forall{m}$ since we work at half-filling and the Hamiltonian~\eqref{eq:hamiltonian_bosons} has a particle-hole symmetry, resulting in $\bigl\langle\hat{n}_m\bigr\rangle=1/2~\forall{m}$. Because $\mathcal{N}_m$ is already zero in absence of noise, we do not expect noise to affects its value. On the other hand, for a finite disorder strength ($\mu>0$), the quantity $\overline{\mathcal{N}_m}$ will be finite, and we can study its dependence as a function of the noise strength $p$.

Results in one dimension for $\mu=5$ are shown in Fig.~\ref{fig:noisy_physics_local_density}(a) and show a behavior of the form,
\begin{equation}
    \overline{\mathcal{N}_m\bigl(N,p\bigr)}=\overline{\mathcal{N}_m\bigl(N,p=0\bigr)}\times\mathcal{G}\bigl(pN\bigr),
    \label{eq:noise_scaling_loc_density}
\end{equation}
with $G(X)$ a scaling function. For $X\equiv pN\ll 1$, we find that $G(X)\simeq\exp(-\tilde{\epsilon}X)$ with $\tilde{\epsilon}=15.7(2)$ a fitting parameter, in line with the expectations for a local observable. Here, $N$ has to be interpreted as the circuit depth $D$ following the free-fermionic compression.

In two dimensions, we plot data for $N=4\times 4$ in Fig.~\ref{fig:noisy_physics_local_density}(b). Similar to one dimension, the noise suppresses exponentially the finite value of the observable of Eq.~\eqref{eq:local_density}: As a function of the noise strength $p$, we find $\simeq\exp(-\tilde{\epsilon}p)$ for $p\ll 1$. Here, the circuit depth is included in the fitting parameter $\tilde{\epsilon}=30(2)\times 10^3$, which explains why it is orders of magnitude larger than in one dimension. This highlights the power of the compression scheme in one-dimension to limit the effect of noise by reducing the circuit depth to $O(N)$. From general grounds, we expect Eq.~\eqref{eq:noise_scaling_loc_density} to remain valid in two dimensions with the modification $N\to D$, where $D$ is the circuit depth.

\subsubsection{The coherent density of bosons}

\begin{figure}[t]
    \includegraphics[width=1\columnwidth]{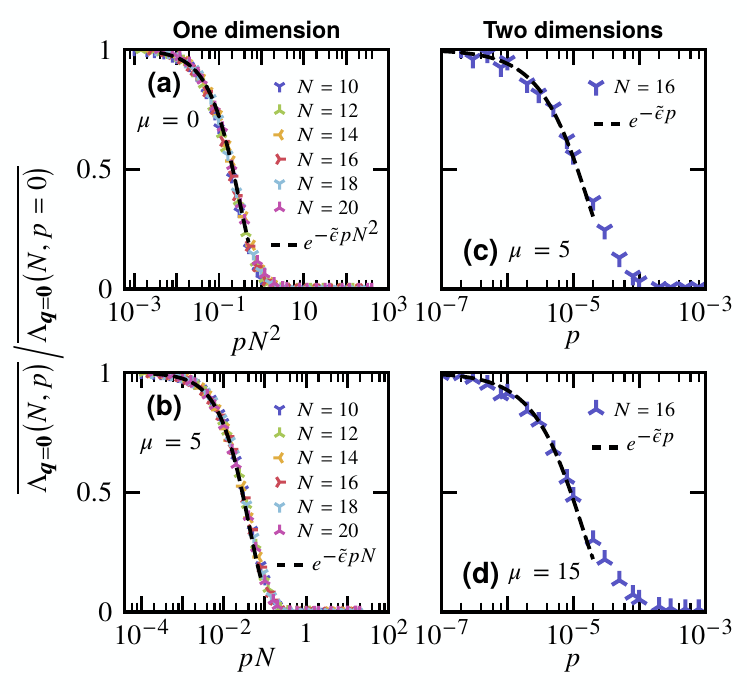} 
    \caption{Coherent density based on the momentum distribution function of Eq.~\eqref{eq:bec_density} at $\boldsymbol{q}=\boldsymbol{0}$. Data normalized by the noiseless case with $p=0$. Left column: One-dimensional case where each data point is averaged over $N_\mathrm{samples}=2^{10}$ random disordered samples, and each of those obtained from $N_\mathrm{shots}=2^{10}$ measurements from noisy simulations. Error bars are smaller than the symbols and not shown. Right column: Two-dimensional case case where each data point is averaged over a hundred random disordered samples, and each of those obtained from $N_\mathrm{shots}=10^2$ measurements from noisy simulations. (a) Data at fixed disorder strength $\mu=0$. For $pN^2\ll 1$, we observe a behavior of the form $\sim\exp(-\tilde{\epsilon}pN^2)$ with $\tilde{\epsilon}=3.28(5)$ a fitting parameter. Here, one of the factors $N$ has to be interpreted as the depth $D$ of the circuit, which scales linearly with $N$ after the free-fermionic compression (b) Data at fixed disorder strength $\mu=5$. For $pN\ll 1$ we observe a behavior of the form $\sim\exp(-\tilde{\epsilon}pN)$ with $\tilde{\epsilon}=24.5(3)$ a fitting parameter. Here, $N$ has to be interpreted as the depth $D$ of the circuit, which scales linearly with $N$ after the free-fermionic compression. (c) Data at fixed disorder strength $\mu=5$ in the delocalized phase (d) Data at fixed disorder strength $\mu=15$ in the localized phase. In both cases, for $p\to 0$, we observe an exponential decay $\sim\exp(-\tilde{\epsilon}p)$ with $\tilde{\epsilon}=60(3)\times 10^3$ ($\mu=5$) and $\tilde{\epsilon}=75(2)\times 10^3$ ($\mu=15$), a fitting parameter.}
    \label{fig:noisy_physics_coherent_density}
\end{figure}

We now turn our attention to the momentum distribution function of Eq.~\eqref{eq:bec_density} at $\boldsymbol{q}=\boldsymbol{0}$. In one dimension, we consider data at criticality ($\mu=0$) and deep in the disordered phase ($\mu=5$). A scaling law involving the noise strength $p$ and the system size $N$ can be obtained, albeit with different scaling variables in the delocalized and localized phases, see Figs.~\ref{fig:noisy_physics_coherent_density}(a) and~\ref{fig:noisy_physics_coherent_density}(b). In particular, we find,
\begin{equation}
    \frac{\overline{\Lambda_{\boldsymbol{q}=\boldsymbol{0}}\bigl(N,p\bigr)}}{\overline{\Lambda_{\boldsymbol{q}=\boldsymbol{0}}\bigl(N,p=0\bigr)}} = \left\{
    \begin{array}{rl}
        \mathcal{G}\bigl(pN^2\bigr)&~~\mathrm{Delocalized},\\
        \\
        \mathcal{G}\bigl(pN\bigr)&~~\mathrm{Localized},
    \end{array}
    \right.
    \label{eq:noise_scaling_bec_density}
\end{equation}
with $\mathcal{G}(X)$ a scaling function. For $X\ll 1$, we find that $G(X)\simeq\exp(-\tilde{\epsilon}X)$ with $\tilde{\epsilon}=3.28(5)$ ($\mu=0$) and $\tilde{\epsilon}=24.5(3)$ ($\mu=5$), a fitting parameter. In the localized scaling, $N$ has to be interpreted as the circuit depth and in the delocalized scaling, one of the factor $N$ has to be interpreted as such. In one dimension, the circuit depth is linear with $N$ thanks to the free-fermionic compression of the circuit. In absence of noise, for a critical behavior, Ref.~\onlinecite{PhysRevB.106.L041109} found a similar to scaling to Eq.~\eqref{eq:noise_scaling_bec_density} for a two-point correlator $C(x,p)$. Since $\Lambda_{\boldsymbol{q}=\boldsymbol{0}}$ is a sum of two-point correlators over all distances $x$, our result fits previous findings. In the localized case, individual correlators can be thought of as local observables extending in real space over a scale corresponding to the localization length $\xi$. Hence, for system sizes such that $N\gg\xi$, the coherent density $\Lambda_{\boldsymbol{q}=\boldsymbol{0}}$ should behave as a local observable with a scaling only involving the circuit depth, fitting the data of Fig.~\ref{fig:noisy_physics_coherent_density}(b) and Eq.~\eqref{eq:noise_scaling_bec_density}. We believe the two behaviors of Eq.~\eqref{eq:noise_scaling_bec_density} can be reconciled by a single scaling function $\mathcal{G}(pN\xi)$ where $\xi$ is substituted by the length of the system for $N\lesssim\xi$. Hence, for $N\gg\xi$ and $\mu\to 0$ where we expect $\xi\sim\mu^{-2}$, one would get a scaling function $\mathcal{G}(pN\mu^{-2})$. However, this regime is out of reach of our simulations for testing.

For completeness, we also perform simulations in two dimensions for $N=4\times 4$ and $\mu=5$, $15$, see Figs.~\ref{fig:noisy_physics_coherent_density}(c) and~\ref{fig:noisy_physics_coherent_density}(d). While the lack of multiple system sizes does not permit to draw conclusions on the $N$-dependence of the noise-induced physics, we find that, similar to one dimension, the coherent density is exponentially suppressed as a function of the noise strength. Although we cannot verify it, we expect Eq.~\eqref{eq:noise_scaling_bec_density} to remain valid in two dimensions, as it was derived from general arguments (one of the parameters $N$ would need to be replaced explicitly by the circuit depth $D$).

\section{Conclusion}
\label{sec:conclusion}
\subsection{Summary}
In this work, we considered the dirty boson problem in one and two dimensions using quantum computing techniques and a programmable quantum computer. In all cases, adiabatic state preparation was used to prepare the ground state of hard-core bosons hopping on a lattice, subject to an on-site random chemical potential of tunable strength.

In one dimension, the model maps to a free-fermionic system, and we leveraged this property to compress the depth of the circuits for the adiabatic state preparation such that it is of the order of the lattice size~\cite{bassman2022constant,Kokcu2021,PhysRevA.105.032420,Camps2022} (as opposed to of the order of evolution time). Upon compression, the circuits were suitable for the current generation of noisy quantum devices. We obtained experimental results on a superconducting quantum computer for $L=6$ qubits. We considered the probability distribution of the local density of bosons displaying a distinctive U-shape at strong disorder. We also studied an off-diagonal two-point correlation function and performed a scaling analysis together with a data collapse to extract critical properties of the transition. We estimated the correlation length exponent $\nu\approx 1.86$, in good agreement with the expected theoretical value of $\nu=2$~\cite{Thouless1972,Giamarchi1987,PhysRevB.37.325}.

In two dimensions, the model does not map to free fermions and the depth of the circuits cannot be compressed anymore. As the depth is directly proportional to the duration of the adiabatic time evolution protocol, the circuits are too large to be reliably executed on the current generation of quantum hardware. Instead, we performed large-scale classical state vector simulations on graphics processing units~\cite{cuquantum}, which emulate the performance of a quantum computer. Our results from emulation of an ideal (i.e., noise-free) quantum computer validated our algorithmic approach to obtaining the ground state physics of two-dimensional disordered systems, demonstrating promising prospects for the next generation of quantum computers with more qubits of better quality.

Finally, we simulated dirty bosons via emulation of a noisy quantum computer, based on a depolarizing channel of Pauli operators with a parameter $p$ controlling the strength of the noise. We sought to understand how noise affects the physical properties of the final quantum state. In the noiseless limit ($p\to 0$), we found that an observable $O$ is generically modified by the noise as $O(p)\sim O(p=0)\times{e}^{-\epsilon p}$ with $\epsilon$ a parameter. In one dimension, we performed a systematic analysis with the system size $N$ and the circuit depth, scaling as $N$. We found that the noise-induced physics is described by scaling relation $O(p)=O(p=0)\times\mathcal{G}(X)$ with $G(X)\simeq{e}^{-\epsilon X}$ for $X\ll 1$. For local observables, such as the local density, we have $X\equiv pN$, while for global quantities, such as the coherent density of bosons, we have $X\equiv pN^2$ and $X\equiv pN$ in the localized and delocalized phases, respectively. The different behaviors are understood through the existence of the localization length $\xi$ making non-local physical observables (e.g., a correlation function) effectively local over this length scale as long as $\xi\ll{N}$. We also advanced a scenario reconciling the two regimes as one tunes the delocalization-to-localization transition. We believe that understanding how noise modifies the genuine properties of different phases is fundamental for leveraging noisy intermediate-scale quantum devices for condensed matter, extending the work of Ref.~\onlinecite{PhysRevB.106.L041109}.

\subsection{Perspectives}

For an adiabatic state preparation protocol to be successful, its total duration $T$ should scale as $T\gtrsim\Delta^{-2}_\mathrm{min}$ with $\Delta_\mathrm{min}$ the minimum spectral gap between the ground state and the first excited state of the interpolating Hamiltonian~\cite{RevModPhys.90.015002}. In the thermodynamic limit, the Bose glass phase is gapless, meaning that the gap at the end of the interpolation goes as $\Delta\to 0$. However, on a finite system, the gap will be finite, but can be exponentially small for some disorder configurations (we studied the distribution of the gap in the main text); this is a general feature of glassy systems~\cite{BAPST2013127}. The existence of arbitrarily small gaps is also an issue in classical methods like quantum Monte Carlo: it works at a finite temperature $\Theta$---the algorithm's complexity scales linearly with $\Theta$ in the absence of a sign problem~\cite{sandvik2010}---which needs to be much smaller than $\Delta$ in order to probe ground state properties. Nevertheless quantum Monte Carlo is still the method of choice to simulate thousands of interacting particles~\cite{alvarez2015,ng2015,doggen2017}, meaning that not perfectly capturing the ground state of configurations with small spectral gaps is not so bad when considering a statistical ensemble over random disordered realizations. Whether this is the case for an adiabatic state preparation remains to be seen. Inspired by the quantum approximate optimization algorithm~\cite{Farhi2014,Farhi2014b,Farhi2016}, one could leave instead the parameters of the evolution gates in Eqs.~\eqref{eq:unitary_z} and~\eqref{eq:unitary_xy} as variational parameters to find an adiabatic shortcut in the state preparation.

In addition to the Bose glass phase and fundamental questions that remain pending on its exact nature (see the discussion in the introduction), one could use a quantum computer to simulate other bosonic phases of matter which are hard to simulate classically due to the existence of a sign problem in quantum Monte Carlo simulations. For instance, one could study Bose metals~\cite{PhysRevLett.106.046402,PhysRevLett.107.077201} or a superglass phase~\cite{PhysRevB.105.174203,PhysRevB.85.104205}---though frustration might not be a necessary ingredient in the latter~\cite{PhysRevLett.116.135303}. In particular, we believe our results on the noise-induced physics will prove useful for such future studies, once more and better qubits become available.

\begin{acknowledgments}
    M.D. gratefully acknowledges D. Azses, E. G. Dalla Torre, B. Evert, J. E. Moore, and M. J. Reagor for recent collaborations on related works. L.B.O., R.V.B., D.C., and W.A.d.J. were supported by the Office of Science, Office of Advanced Scientific Computing Research Accelerated Research for Quantum Computing Program of the U.S. Department of Energy under Contract No. DE-AC02-05CH11231. This research used the Lawrencium computational cluster resource provided by the IT Division at the Lawrence Berkeley National Laboratory and resources of the National Energy Research Scientific Computing Center (NERSC), a U.S. Department of Energy Office of Science User Facility located at Lawrence Berkeley National Laboratory, operated under Contract No. DE-AC02-05CH11231 using NERSC award DDR-ERCAP0022246 under the QIS@Perlmutter program.
\end{acknowledgments}

\bibliography{references}

\end{document}